\documentclass[aps,showpacs,floatfix,preprint,pre,groupedaddress]{revtex4}
\usepackage{times}
\usepackage{xspace}
\usepackage{graphicx,morefloats}
\usepackage{color}
\usepackage{amsmath, amsthm, amssymb}
\usepackage{enumerate}

\begin{document}
\title{The effects of psammophilous plants on sand dune dynamics}

\author{Golan Bel}
\email{bel@bgu.ac.il}
\author{Yosef Ashkenazy}
\email{ashkena@bgu.ac.il}
\affiliation{Department of Environmental Physics, Blaustein Institutes for Desert Research, Ben-Gurion University of the Negev, Sede Boqer Campus 84990, Israel}

\date{\today}

\begin{abstract}
  Psammophilous plants are special plants that flourish in sand moving
  environments. There are two main mechanisms by which the wind affects
  these plants: (i) sand drift exposes roots and covers
  branches--the exposed roots turn into new plants and the covered
  branches turn into new roots; both mechanisms result in an enhanced
  growth rate of the psammophilous plant cover of the dunes; (ii)
  strong winds, often associated with sand movement, tear branches and
  seed them in nearby locations, resulting in new plants and an enhanced
  growth rate of the psammophilous plant cover of the dunes.
Despite their important role in dune dynamics, to our knowledge, psammophilous plants have never been incorporated into
mathematical models of sand dunes. Here, we attempt to model the
  effects of these plants on sand dune dynamics. We construct a set of
  three ordinary differential equations for the fractions of surface
  cover of regular vegetation, biogenic soil crust and psammophilous
  plants. The latter reach their optimal growth under (i) specific
  sand drift or (ii) specific wind power. We show that psammophilous
  plants enrich the sand dune dynamics. Depending on the
  climatological conditions, it is possible to obtain one, two, or three
  steady dune states.  The activity of the dunes can be associated
  with the surface cover--bare dunes are active, and dunes with
  significant cover of vegetation, biogenic soil crust, or
  psammophilous plants are fixed. Our model shows that under suitable
  precipitation rates and wind power, the dynamics of the different
  cover types is in accordance with the common view that
  dunes are initially stabilized by psammophilous plants that reduce sand
  activity, thus enhancing the growth of regular vegetation that
  eventually dominates the cover of the dunes and determines their
  activity.
\end{abstract}


\maketitle

\section{Introduction}{\label{intro}}

Sand dunes cover vast areas in arid and coastal regions \cite[$\sim
10\%$,][]{Pye-1982:Morpholgical, Pye-Tsoar-1990:aeolian,
  Thomas-Wiggs-2008:aeolian} and are considered to be an important
component of geomorphological \cite[][]{Bagnold-1941:physics} and
ecological \cite[][]{Tsoar-2008:land,
  Veste-Littmann-Breckle-Yair-2001:role} systems. On one hand, active
sand dunes are a threat to humans
\cite[][]{Dong-Chen-He-et-al-2004:controlling,
  Khalaf-Al-Ajmi-1993:aeolian}, while, on the other hand, they are
associated with unique ecosystems that increase biodiversity
\cite[][]{Shanas-Abu-Galyun-Alshamlih-et-al-2006:reptile} and
thus are important to humans. Human activities can affect sand dune
ecosystems \cite[][]{Tsoar-2008:land,
  Veste-Littmann-Breckle-Yair-2001:role}. Sand dunes may be sensitive
to climate change
\cite[][]{Thomas-Knight-Wiggs-2005:remobilization,
  Ashkenazy-Yizhaq-Tsoar-2012:sand}, and it has been claimed that they influence the
climate system through changes in their albedo
\cite[][]{Otterman-1974:baring, Charney-Stone-Quirk-1975:drought}.

The wind is the main driving force of sand dunes
\cite[][]{Tsoar-2005:sand}. The migration rate of sand dunes  is proportional
to the wind power, which is a non-linear function of the wind speed
\cite[][]{Fryberger-1979:dune}. Thus, dunes mainly migrate during a
small number of extreme wind events. Dunes may be stabilized by vegetation
and/or biogenic soil crust (BSC)
\cite[][]{Danin-Bar-Or-Dor-Yisraeli-1989:role}; since vegetation can
only exist above a certain precipitation threshold \cite[typically
$\sim 50 mm/yr$,][]{Tsoar-2005:sand}, sand dune dynamics and activity 
in arid regions are strongly affected by the precipitation rate.

{Many experimental \cite[][]{Bagnold-1941:physics,
    Fryberger-1979:dune, Pye-Tsoar-1990:aeolian} and theoretical
  \cite[][]{Bagnold-1941:physics,
    Andreotti-Claudin-Douady-2002:two-dimensional,
    Duran-Herrmann-2006:vegetation,
    Luna-Parteli-Duran-Herrmann-2009:modeling,
    Reitz-Jerolmack-Ewing-Martin-2010:barchan,
    Nield-Baas-2008:influence,
    Kok-Parteli-Michaels-Karam-2012:physics} works have been devoted to
  uncovering the mechanisms behind the geomorphology of sand dunes.} Most
of these models focused on the dune patterns and their corresponding
scaling laws, on dune formation, and on the transition from one type
of dune to another. These mathematical/ physical models usually
require a long integration time, therefore only enabling the
simulations of relatively small dune fields. An alternative approach is
to model the vegetation and BSC cover of the dunes, ignoring dune
patterns and 3D dune dynamics, and to determine dune stability (active
or fixed) according to the fraction of cover of vegetation and
BSC; bare dunes are active, while vegetated and/or BSC covered dunes
are fixed \cite[][]{Yizhaq-Ashkenazy-Tsoar-2007:why,
  Yizhaq-Ashkenazy-Tsoar-2009:sand,
  Kinast-Meron-Yizhaq-Ashkenazy-2013:biogenic,
  Yizhaq-Ashkenazy-Levin-Tsoar-2013:spatiotemporal}. Such models
require a relatively short computation time and have been used to explain
the bi-stability of active and fixed dunes under similar climatic
conditions. In addition, it is possible to model the development of a
2D vegetation cover by considering the spatial effect of the wind and
the diffusion of vegetation
\cite[][]{Yizhaq-Ashkenazy-Levin-Tsoar-2013:spatiotemporal}. Both
  observations \cite[][]{Veste-Littmann-Breckle-Yair-2001:role}
  and models \cite[][]{Kinast-Meron-Yizhaq-Ashkenazy-2013:biogenic}
  indicate that BSC plays an important role in dune stabilization in
  arid regions with relatively weak winds.

The movement of windblown sand is a stress to ``regular'' vegetation
(hereafter ``vegetation'').  Some species have evolved to tolerate,
and even flourish in, moving-sand environments.  These plants are
are called ``psammophilous plants'' \cite{Danin-1991:plant,
  Danin-1996:plants}. Psammophilous plants have developed several
physiological mechanisms to survive and benefit from sand
drift. Here, we focus on the following interactions of these plants
with sand drift: i) exposure of roots due to sand movement; some of
these plants can grow leaves on the exposed roots, thereby increasing
their photosynthesis and their growth rates; ii) burial of branches by
the windblown sand; some of these plants are able to use the buried
branches as roots, thereby enhancing the growth rate of aboveground biomass
without changing the root:shoot ratio; iii) tearing of branches/leaves
by the wind and their burial by the sand; in some of these plants, this
is a mechanism that enhances the clonal growth through the development of
new plants from the buried branches. These interactions may be divided
into two groups: interactions (i) and (ii) whose rate of occurrence
and efficiency are determined by the actual sand drift (hereafter, we
will refer to this group as mechanism I), and interaction (iii) whose
rate and efficiency are determined by the wind drift potential
(hereafter, mechanism II). We note that this is an oversimplified
classification of the interactions of psammophilous plants with the
wind and the sand drift.

Psammophilous plants play an important role in dune stabilization. Due
to their adaptation to sand moving environments, they are the first to
develop (under suitable environmental conditions) in bare and active
sand dunes \cite[][]{Danin-1996:plants}. Once sufficiently dense
psammophilous plant cover is established, the sand movement is
reduced accordingly, enabling the development of vegetation and
BSC. This development further reduces the sand activity, suppressing
the growth of psammophilous plants, and further enhancing the
growth of vegetation and BSC.  This process may continue until the
dunes become fixed and reach a steady state associated with the
environmental conditions. Despite their important role in dune
stabilization, to our knowledge, psammophilous plants have never been incorporated into
mathematical models of sand dunes.

The major goal of this study is to investigate the dynamics of
psammophilous plants on sand dunes when coupled to vegetation and BSC
dynamics. The model suggested below is a natural extension of the
model of \cite{Yizhaq-Ashkenazy-Tsoar-2007:why} and others
\cite[][]{Yizhaq-Ashkenazy-Tsoar-2009:sand,
  Kinast-Meron-Yizhaq-Ashkenazy-2013:biogenic,
  Yizhaq-Ashkenazy-Tsoar-2009:sand}. The model describes the
development of vegetation, BSC, and psammophilous plants on sand
dunes, taking into account the effects of the wind and the
precipitation. We suggest two ways to model psammophilous
plants. The first approach aims to describe ``mechanism I,'' in which the
growth of the psammophilous plants is optimal under a specified sand
flux. The second approach describes ``mechanism II,'' in which
psammophilous plants reach their optimal growth under a specified
optimal wind power (or drift potential, defined below). The setup up of the models of mechanisms I and II is different, since the drift potential, used to model the optimal growth due to mechanism II, is not affected by the
actual dune cover, while the sand flux that is used to model mechanism
I is strongly affected by the dune cover. The modeling of mechanism
I yielded a richer bifurcation diagram (steady states map) compared to
the modeling of mechanism II. Both modeling approaches show that for some climatic conditions (a region in the drift potential and precipitation rate parameter space), the psammophilous
plants act as pioneers in colonizing sand dunes, followed by
vegetation and/or BSC that dominates the sand dune cover toward its
stabilization. This dynamics is in agreement with the scenario
suggested by \cite{Danin-1996:plants}.

\section{The model}
Our model for psammophilous plants (coupled to vegetation and BSC)
follows previously suggested mean field models
\cite{Yizhaq-Ashkenazy-Tsoar-2007:why,Yizhaq-Ashkenazy-Tsoar-2009:sand,Kinast-Meron-Yizhaq-Ashkenazy-2013:biogenic}
for the dynamics of vegetation and BSC cover of sand dunes.  The
dynamical variables in our model are the fractions of regular
vegetation cover, $v$, BSC cover, $b$, and psammophilous plant cover,
$v_p$, where $v_p$ is a new variable added to the model described in
\cite{Kinast-Meron-Yizhaq-Ashkenazy-2013:biogenic}.

The effects considered in the previous models
\cite{Yizhaq-Ashkenazy-Tsoar-2007:why,Yizhaq-Ashkenazy-Tsoar-2009:sand,Kinast-Meron-Yizhaq-Ashkenazy-2013:biogenic}, as well as in this model, may
be divided into three categories: effects that are not related to the
wind, effects that are directly related to the wind, and effects that
are indirectly related to the wind (representing aeolian effects).  
The effects that are not related to the wind include 
the growth and mortality of the different cover types. We assume a
logistic type growth
\cite[][]{Baudena-Boni-Ferraris-Hardenberg-Provenzale-2007:vegetation}. The
natural growth rate, $\alpha_{j}\left(p\right)$ ($j$ stands for the
cover type, either $b$, $v$ or $v_p$), depends on the precipitation
rate, $p$; for simplicity and consistency with previous works, we adopt the form of
\cite{Yizhaq-Ashkenazy-Tsoar-2007:why,Yizhaq-Ashkenazy-Tsoar-2009:sand,Kinast-Meron-Yizhaq-Ashkenazy-2013:biogenic,Yizhaq-Ashkenazy-Levin-Tsoar-2013:spatiotemporal},
 \begin{align}
   \alpha_j\left(p\right)\equiv
   \alpha\_max_j\left(1-\exp\left(\frac{p-p\_min_j}{c_j}\right)\right)
   \ \ \ \ \ \ \ \ j \in \left\{ v,v_p,b \right\}. \label{alphadef}
 \end{align}
 $\alpha\_max_j$ is the maximal growth rate of the $j$'th cover
 type. This maximal growth rate is achieved when the precipitation
 rate, $p$, is high enough not to be a growth limiting
 factor and when the other climatic conditions are optimal.  In
 addition, we consider the spontaneous growth of the cover types 
 (growth occurring even in bare dunes) due to effects not modeled 
 here, such as the soil seed bank, underground roots and seed
 dispersal by the wind and animals. These effects are characterized 
 by spontaneous growth rates, $\eta_j$.
 The wind-independent mortality is accounted for by an effective
 mortality rate for each cover type, $\mu_{j}$.

 In modeling the direct and indirect effects of the wind, we use the
 wind drift potential, $D_p$, as a measure of the wind power
 \cite[][]{Fryberger-1979:dune}; $D_p$ is linearly proportional to the sand
 drift.
 The wind drift potential is defined as
\begin{align}
	D_p\equiv \langle U^2\left(U-U_t\right)\rangle, \label{DPdef}
\end{align}
where $U$ is the wind speed (at $10 m$ height above the ground)
measured in knots ($1 knot=0.514 m/s$), $U_t=12 knots$ is the threshold
wind speed necessary for sand transport, and the $\langle \cdot\rangle$
denotes a time average. When the wind speed, $U$, is measured in
knots, $D_p$, is measured in vector units, $VU$. $D_p$ provides only
the potential value of sand drift; in the case of unidirectional wind,
it coincides with the resultant wind drift potential (RDP), which also takes
into account the wind direction. Here, we assume that the winds
are unidirectional and use $D_p$ instead of RDP.

Two important, direct and indirect, wind effects are considered in our
model. The direct damage/mortality by the wind is proportional to the
square of the wind speed (which is proportional to the wind
stress). For simplicity, and in order to minimize the number of the
parameters in the model, we assume that the direct damage by the wind
is proportional to $D_p^{2/3}$. The indirect wind effect is the
movement of windblown sand—that is, sand drift.  The sand drift is equal to
the drift potential multiplied by the amount of sand multiplied by a
function, $g(v,v_p)$, which accounts for the sand-drift shading by the
vegetation \cite[][]{Lee-Soliman-1977:investigation,
  Wolfe-Nickling-1993:protective}. The sand-drift shading function is
assumed to be a step-like function that, above some critical value of
the vegetation cover, $v_c$, drops to zero, while for values of the
vegetation cover much lower than $v_c$, it obtains its maximal value,
$1$ \cite[][]{Lee-Soliman-1977:investigation}. For simplicity, it is
assumed that $g(v,v_p)$ is a function of the difference between the
actual fraction of vegetation cover, $v+v_p$, and the critical value
$v_c$. In the previous models
\cite{Yizhaq-Ashkenazy-Tsoar-2007:why,Yizhaq-Ashkenazy-Tsoar-2009:sand,Kinast-Meron-Yizhaq-Ashkenazy-2013:biogenic,Yizhaq-Ashkenazy-Levin-Tsoar-2013:spatiotemporal},
the sand drift was considered as a damaging effect, increasing the
mortality of regular vegetation and BSC due to root exposure and aboveground biomass burial by the sand.

Here, we focus on the role of psammophilous plants, and therefore,
we consider their cover fraction, $v_p$, as an additional dynamical
variable with a unique interaction with the sand drift. The
psammophilous plants reach their maximal growth rate under optimal
sand drift \cite{Danin-1996:plants}, providing them with the necessary
rate of sand cover and/or exposure and branch/leaf seeding (the two
mechanisms described in the introduction). These unusual optimal
conditions yield a different dynamics of psammophilous
plants. This dynamics, when coupled with the dynamics of the regular
vegetation and BSC, leads to interesting and complex bifurcation
diagrams (steady states) of the sand dunes and their cover types. The
complete set of equations describing the dynamics of the sand dune
cover types is:
\begin{subequations}
\begin{align}
	\partial_t v&=\alpha_v\left(p\right) \left(v+\eta_v\right) s-\gamma_v D_p^{2/3} v-\epsilon_v v D-\mu_v v, \label{dynv} \\
 	\partial_t v_p&=\alpha_{v_p}\left(p\right)\left(v_p+\eta_{v_p}\right)s-\gamma_{v_p}D_p^{2/3}v_p-\epsilon_{v_p} v_p f_i\left(D_p,v,b,v_p\right)-\mu_{v_p} v_p, \label{dynvp} \\
 	\partial_t b&=\alpha_{b}\left(p\right)\left(b+\eta_b\right)s-\epsilon_{b} b D-\mu_b b. \label{dynb}
\end{align}
\end{subequations}
We introduce the following notations:
$s$ is the fraction of bare sand
\begin{align}
 s\equiv 1-v-v_p-b. \label{sdef}
\end{align}
The sand drift, $D$, is defined as: 
 \begin{align}
  D\equiv D_p \times g\left(v+v_p-v_c\right)\times s. \label{Ddef}
 \end{align}
The sand drift shading function is defined as:
\begin{align}\label{gtdef}
g\left(x\right)\equiv \left\{
\begin{array}{l l}
1  & \quad x<-1/d \\
0.5\left(1-xd\right) & \quad -1/d<x<1/d \\
0 & \quad x>1/d
\end{array}
\right.
\end{align}
where the parameter $d$ determines the sharpness of the transition
from total sand-drift shading to its absence.

The effect of sand drift on psammophilous plants is different than
its effect on the other types of sand cover. Here, we consider two
different options of modeling this unique interaction of
psammophilous plants with sand drift, mechanisms I and II which
were explained above. These two mechanisms are modeled using different
forms of the function $f_i\left(D_p,v,b,v_p\right)$.

In the first approach (mechanism I), only the sand drift is assumed to
affect the dynamics of $v_p$. Therefore,
$f_I\left(D_p,v,b,v_p\right)=Q\left(D\right)$. $Q\left(D\right)$
obtains its minimal value, $Q\left(D_{opt}\right)=0$, for
$D=D_{opt}$. Away from the optimal sand drift conditions,
$Q\left(D\right)$ is larger than zero and hence introduces a mortality
term due to the non-optimal sand drift conditions.  This form of the
function $Q\left(D\right)$ reflects the fact that the maximal growth
rate, $\alpha\_max_{v_p}$, is assumed to be under optimal sand drift
conditions.  We thus used the following form of $Q\left(D\right)$
\begin{align}\label{QDdef}
Q\left(D\right)\equiv \left\{
\begin{array}{l l}
 \frac{1}{\sigma^2}\left(D-D_{opt}\right)^2 & \quad \left|D-D_{opt}\right|<\sigma \\
1 & \quad \left|D-D_{opt}\right|>\sigma 
\end{array}
\right.
\end{align}
This choice of the function reflects the behavior described
above. Other choices of the function $Q\left(D\right)$ (such as
$1-\exp\left(\left(D-D_{opt}\right)^2/\sigma^2\right)$) yielded
similar results. Therefore, we decided to focus on this specific,
rather simple, choice. It is important to note that this form of
the function ensures that $v_p$ is never high enough (compared to
$v_c$) to create a total sand-drift shading (this would result in a
vanishing sand drift and therefore, in far from optimal conditions).

The second approach to model the enhanced growth of psammophilous
plants under sand drift conditions (mechanism II) is simpler and
assumes that the optimal growth conditions are achieved under an
optimal wind drift potential rather than under optimal sand drift
conditions. Therefore, we assume that
$f_{II}\left(D_p,v,b,v_p\right)=D\times R\left(D_p\right)$, where
\begin{align}
	R\left(D_p\right)\equiv 1- \rho
        \exp\left(\frac{\left(D_p-{D_{p,{\rm opt}}}\right)^2}{2\sigma^2}\right).
\end{align}
The parameter $\rho>1$ and is set to ensure that the maximal value of
$v_p$ won't exceed $v_c$. Note that when $(D_p-D_{p,{\rm opt}})^2\gg
2\sigma^2$, this vegetation growth term turns into an indirect
mortality term, similar to the interaction of regular vegetation with the sand drift.

Below, we refer to the different approaches as model I and II,
respectively (corresponding to mechanisms I and II for the enhanced
net growth of psammophilous plants).  For consistency with previous
studies
\cite{Yizhaq-Ashkenazy-Tsoar-2007:why,Yizhaq-Ashkenazy-Tsoar-2009:sand,Kinast-Meron-Yizhaq-Ashkenazy-2013:biogenic},
we use the following values for the
parameters:$\alpha\_max_v=0.15/yr$, $p\_min_{v}=50mm/yr$,
$c_{v}=100mm/yr$, $\eta_{v}=0.2$, $\mu_v=0$,
$\gamma_{v}=0.0008VU^{3/2}/yr$, $\epsilon_{v}=0.001/VU/yr$.
$\alpha\_max_b=0.015/yr$, $p\_min_{b}=20mm/yr$, $c_{b}=50mm/yr$,
$\eta_{b}=0.1$, $\epsilon_{b}=0.0001/VU/yr$.
$\alpha\_max_{v_p}=0.15/yr$, $p\_min_{v_p}=50mm/yr$,
$c_{v_p}=100mm/yr$, $\eta_{v_p}=0.2$, $\sigma=100VU$,
$\gamma_{v_p}=0.0006VU^{3/2}/yr$, $v_c=0.3$, $d=15/2.35\approx6.383$
(for a discussion of the choice of the value of $d$, see
\cite{Yizhaq-Ashkenazy-Tsoar-2009:sand}). In model I,
$\epsilon_{v_p}=0.2/yr$, $\mu_{v_p}=0$ and $D_{opt}=300VU$, while in
model II, $\epsilon_{v_p}=\epsilon_v=0.001/VU/yr$, $\mu_{v_p}=1.2/yr$,
${D_{p,{\rm opt}}}=300VU$, and $\rho=5.0072$. In what follows, the
drift potential, $D_p$, will be measured in units of $VU$, the
precipitation rate, $p$, in units of $mm/yr$ and the mortality rate of
the BSC, $\mu_b$, in units of $1/yr$. Time is measured in years. For
convenience, we drop the units hereafter. Justification regarding the
choice of the parameters and the model setup can be obtained from
\cite[][]{Yizhaq-Ashkenazy-Tsoar-2007:why,
  Yizhaq-Ashkenazy-Tsoar-2009:sand,
  Kinast-Meron-Yizhaq-Ashkenazy-2013:biogenic,
  Yizhaq-Ashkenazy-Levin-Tsoar-2013:spatiotemporal}. Note that for
simplicity, we do not include direct competition terms between the
different models’ variables, unlike
\cite[][]{Kinast-Meron-Yizhaq-Ashkenazy-2013:biogenic}. Below, we
present results for different values of $D_p$, $p$ and $\mu_b$.

\section{Results}
\begin{figure}
\includegraphics[width=\linewidth]{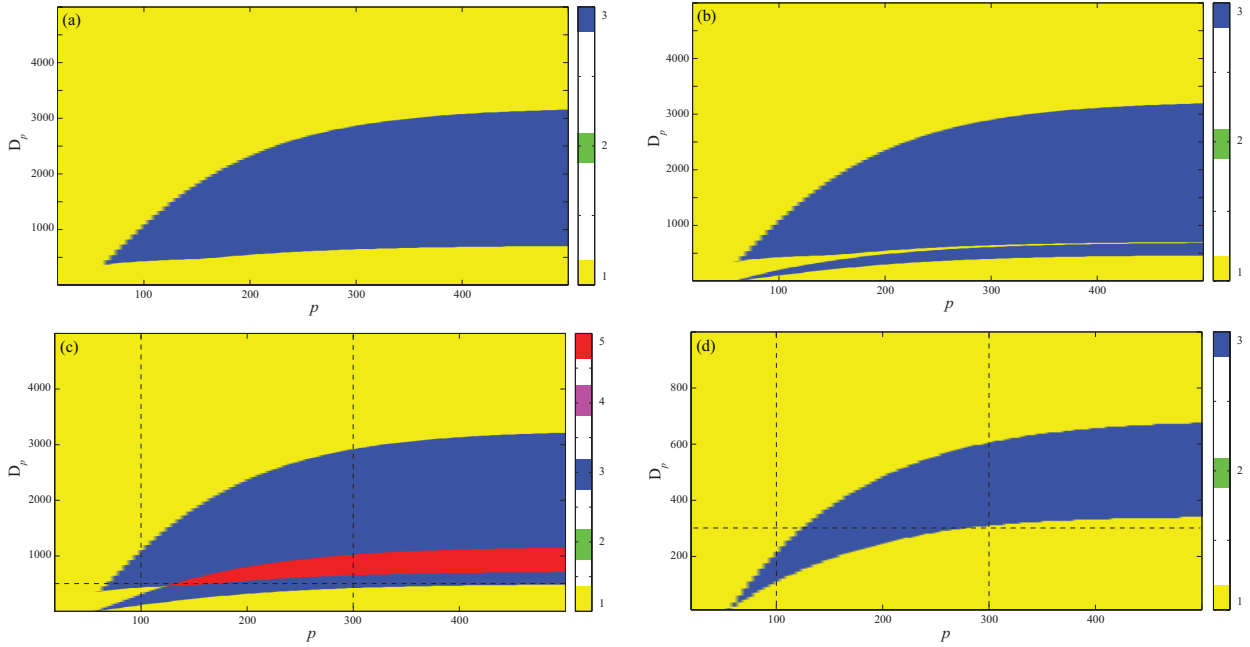}%
\caption{Maps of the number of solutions as a function of the
  precipitation rate, $p$ and the wind drift
  potential $D_p$. Panels (a)-(c) correspond to model I with BSC
  mortality rates $\mu_b=0.001,0.006,0.01$, respectively. Panel (d)
  corresponds to model II and BSC mortality rate $\mu_b=0.006$.
  \label{fig:Nsol}}
\end{figure}

We started our model analysis by studying the number of physical
solutions ($0<v,v_p,b<1$) for given wind conditions characterized by
$D_p$, and precipitation rate, $p$; these are the two main climatic
factors that affect sand dune dynamics. We found that the number of
physical solutions strongly depends on the maximal value of the BSC
cover, which is determined by the BSC mortality rate, $\mu_b$, and the
other parameters. In Fig. \ref{fig:Nsol}, we show maps of the total
number of physical solutions (both stable and unstable) for different
values of $p$ and $D_p$. Panels (a)-(c) correspond to model I and BSC
mortality rates $\mu_b=0.001,0.006,0.01$, respectively. Panel (a)
corresponds to a small value of the BSC mortality rate, $\mu_b=0.001$,
and it shows the existence of a typical bi-stability region (the
region with three solutions, two of which are stable and one is
unstable). Panel (b) corresponds to a higher value of the BSC
mortality rate, $\mu_b=0.006$, for which we have two regions of
bi-stability. Panel (c) corresponds to an even higher value of the BSC
mortality rate, $\mu_b=0.01$, for which we obtain two regions of
bi-stability and, in addition, a region of tri-stability (the total
number of steady states is 5). Panel (d) corresponds to model II with
$\mu_b=0.006$. It shows the existence of a bi-stability range. For much smaller BSC mortality
rates, model II doesn't show a bi-stability region;
higher values of $\mu_b$ change the location (in the parameter space)
of the bi-stability region but do not result in a qualitatively
different bifurcation diagram.

Fig. \ref{fig:Nsol} shows that the two approaches, adopted in models I
and II respectively, result in different numbers of physical steady
state solutions. In model I, for all three values of $\mu_b$
considered here, we have at least one range with three physical
solutions (as shown in Fig. \ref{fig:Nsol}). Two of the three
solutions are stable and one is unstable. As we increase the mortality
rate of the BSC (see Fig. \ref{fig:Nsol}(b)), and therefore, reduce
the maximal value of $b$, a second region of bi-stability
appears. A further increase of $\mu_b$ results in an overlap of the two
bi-stability regions, and therefore, in a range of tri-stability in
which we have five physical solutions (three stable solutions and two
unstable ones, see Fig. \ref{fig:Nsol}(c)). The two bi-stability
regions are due to the different actions of the sand-drift shading on
the regular and the psammophilous plants. Small values of $\mu_b$
allow for high values of $b$, and therefore, the only possible
bi-stability is due to low or high values of $v_p$ which, by shading,
reduces the sand drift even for high values of $D_p$. It is important
to note that in this case, one of the states corresponds to active
dunes, while the other one corresponds to marginally stable dunes. The
psammophilous plants can never cover the dunes to the extent to which
there is no sand drift because they cannot survive away from the
optimal sand drift, ${D_{opt}}$. For higher values of $\mu_b$, the
previously observed bi-stability of active and stable dunes, due to
sand-drift shading by regular plants
\cite{Kinast-Meron-Yizhaq-Ashkenazy-2013:biogenic}, appears and creates
the second range of bi-stability for lower values of $D_p$. Further
increasing the BSC mortality rate results in an overlap of the two
bi-stability ranges, and therefore, in a range of tri-stability. In
model II, there is, at most, one region of bi-stability, as shown in
Fig. \ref{fig:Nsol}(d). For the parameters explored here, we could not
identify two distinct mechanisms of bi-stability. For much
smaller values of $\mu_b$, there is no bi-stability range, and for all
values of $p$ and $D_p$, there is only one physical solution. For
larger values of $mu_b$, the bifurcation diagram is qualitatively the
same as the one presented in Fig. \ref{fig:Nsol}(d). Similar behavior
is obtained when considering only vegetation
\cite[][]{Yizhaq-Ashkenazy-Tsoar-2009:sand} and when considering BSC
in addition to regular vegetation
\cite{Kinast-Meron-Yizhaq-Ashkenazy-2013:biogenic}.

\begin{figure}
\includegraphics[width=\linewidth]{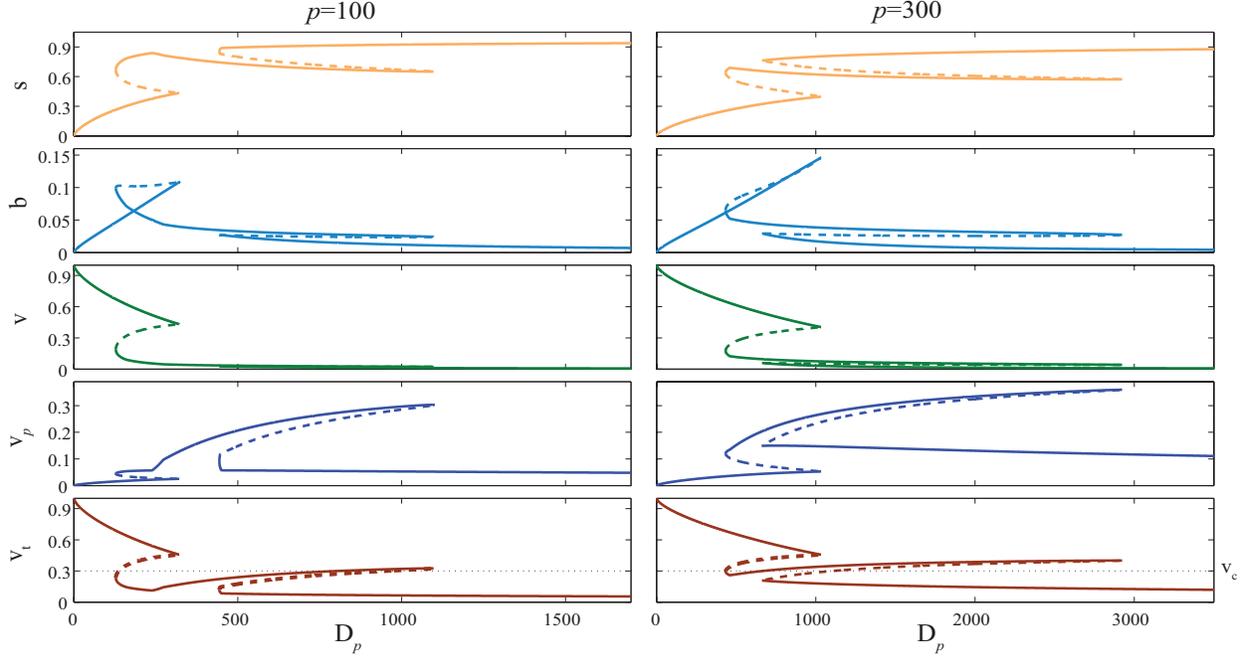}%
\caption{Bifurcation diagrams versus the drift potential, $D_p$, as
  predicted by model I along the vertical dashed lines indicated 
  in Fig. \ref{fig:Nsol}(c). The left column corresponds
  to precipitation rate, $p=100$, and the right column to $p=300$. The
  rows (from top to bottom) correspond to the fractions of uncovered
  sand, $s$, BSC cover, $b$, psammophilous plant cover, $v_p$,
  regular vegetation cover, $v$, and the total sand-drift shading
  vegetation, $v_t\equiv v_p+v$ (the dotted line marks the
  critical value of the vegetation cover for sand-drift shading,
  $v_c$). The solid (dashed) lines correspond to stable (unstable)
  states. The BSC mortality rate is $\mu_{b}=0.01$. 
  \label{fig:dpbifIp100p300}}
\end{figure}

To better understand the complex steady state phase space, we present
in Fig. \ref{fig:dpbifIp100p300} the bifurcation diagrams, as
predicted by model I, against the drift potential. These bifurcation
diagrams show cross-sections along the vertical dashed lines in
Fig. \ref{fig:Nsol}(c). The two columns correspond to two values of
the precipitation rate. The different rows correspond to the different
cover type fractions. We also present: (i) the total vegetation cover
($v+v_p$) which determines the stability of the dunes, and (ii) the
fraction of exposed sand. Obviously, these two variables may be
extracted from $v$, $b$, and $v_p$ and are only shown for clarity.
For the higher value of the precipitation rate, $p=300$, there are two
$D_p$ ranges of a single stable state (for low and high values of
$D_p$). In addition, there are two ranges of bi-stability--one in which
the dunes may be exposed or densely covered and a second range in
which the dunes may be exposed or partially covered ($v_t\sim
v_c$). In between the two bi-stability ranges, there is a range of
$D_p$ for which we have tri-stability. Namely, the dunes may be
exposed, densely covered or partially covered. For the smaller
precipitation rate, $p=100$, the tri-stability range disappears. 
The value of the BSC mortality was set equal to the value used in
Fig. \ref{fig:Nsol}(c) to capture the more complicated bifurcation
diagrams.

\begin{figure}
\includegraphics[width=0.5\linewidth]{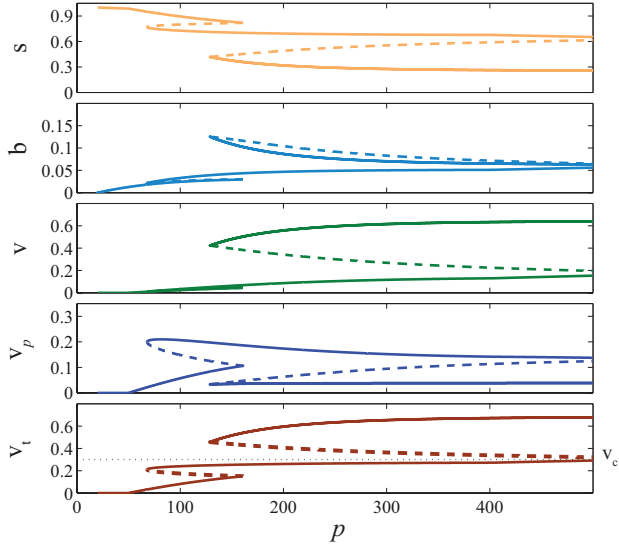}%
\caption{Bifurcation diagrams versus the precipitation rate, $p$, as
  predicted by model I along 
  the horizontal dashed line of Fig. \ref{fig:Nsol}(c). The different rows
  correspond to the cover type fractions as in
  Fig. \ref{fig:dpbifIp100p300}. The drift potential was set to
  $D_p=500$, to capture all the solution branches. The BSC mortality
  rate was set to $\mu_b=0.01$. \label{fig:pbifIdp500}}
\end{figure}

To complete the picture of the bifurcation diagrams, as predicted by
model I, we show in Fig. \ref{fig:pbifIdp500} the bifurcation diagrams
against the precipitation rate for a fixed value of the drift
potential ($D_p=500$, set to ensure that all of the five physical
solutions are captured). These diagrams correspond to a cross-section
along the dashed horizontal line in Fig. \ref{fig:Nsol}(c). In these
diagrams, one can see the onset of bi-stability, followed by the onset
of tri-stability which later on disappears as the precipitation rate
increases.

\begin{figure}
\includegraphics[width=\linewidth]{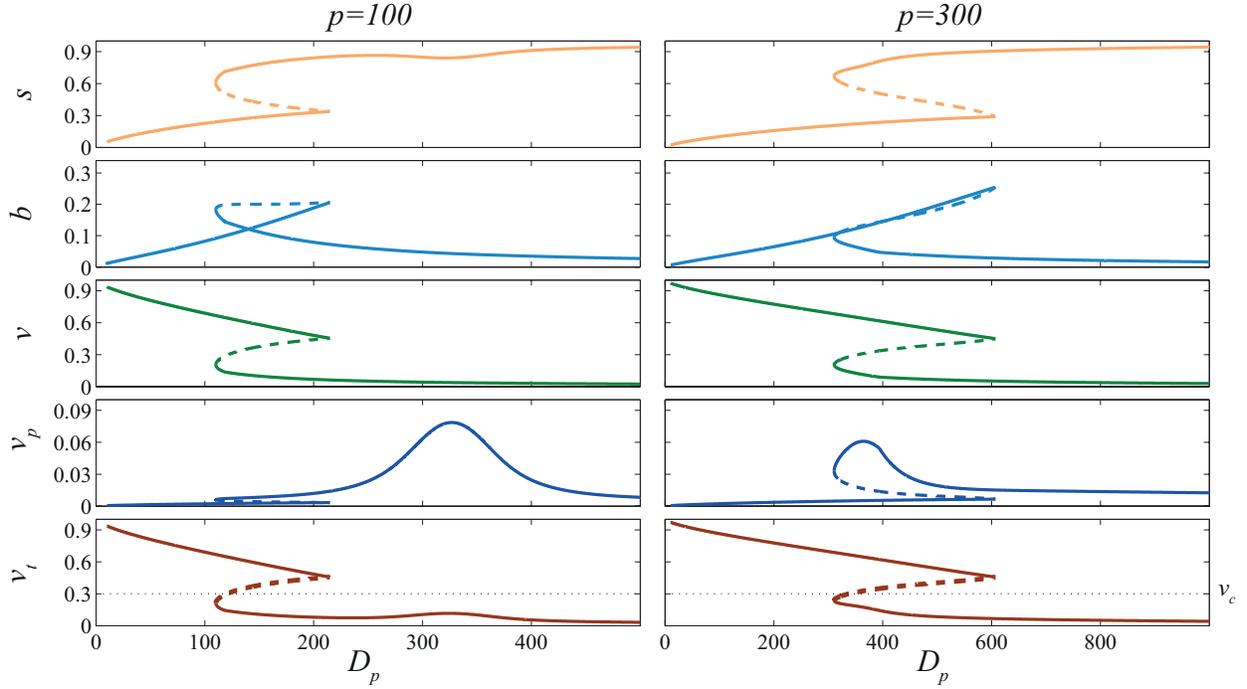}%
\caption{Bifurcation diagrams as predicted by model II along the
  vertical dashed lines of Fig. \ref{fig:Nsol}(d). The
  bifurcation parameter is the drift potential, $D_p$. The different
  rows correspond to the cover type fractions, as in
  Fig. \ref{fig:dpbifIp100p300}. The left column corresponds to
  precipitation rate, $p=100$, and the right column corresponds to
  precipitation rate, $p=300$.
  \label{fig:dpbifII}}
\end{figure}

Fig. \ref{fig:dpbifII} depicts the bifurcation diagrams versus the
drift potential, as predicted by model II, for two values of the
precipitation rate, $p=100$ and $p=300$. 
The bifurcation diagrams are
taken along cross-sections corresponding to the dashed vertical lines
in Fig. \ref{fig:Nsol}(d). For both values of the precipitation rate,
there is only one bi-stability range. However, its width, shape and
location (in the parameter space) are affected by the value of $p$. A
significant difference between model II and model I is the lack in the former of a
steady state corresponding to partially covered dunes ($v_t\sim v_c$).

\begin{figure}
\includegraphics[width=0.5\linewidth]{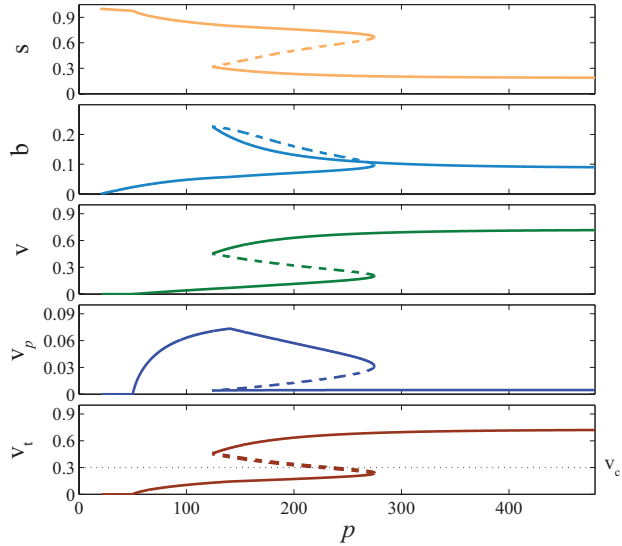}%
\caption{Bifurcation diagrams as predicted by model II. The
  bifurcation parameter is the precipitation rate, $p$. The different
  rows correspond to the cover type fractions. The drift potential was
  set to, $D_p=300$. 
  \label{fig:pbifII}}
\end{figure}

Bifurcation diagrams versus the precipitation rate as predicted by
model II are presented in Fig. \ref{fig:pbifII}. The drift potential
was set to the optimal value for psammophilous plants according to
this model, $D_p=300$ (corresponding to the horizontal dashed line in
Fig. \ref{fig:Nsol}(d)). In these diagrams, one can see the onset of
bi-stability and its disappearance as the precipitation rate
increases.

\begin{figure}
\includegraphics[width=\linewidth]{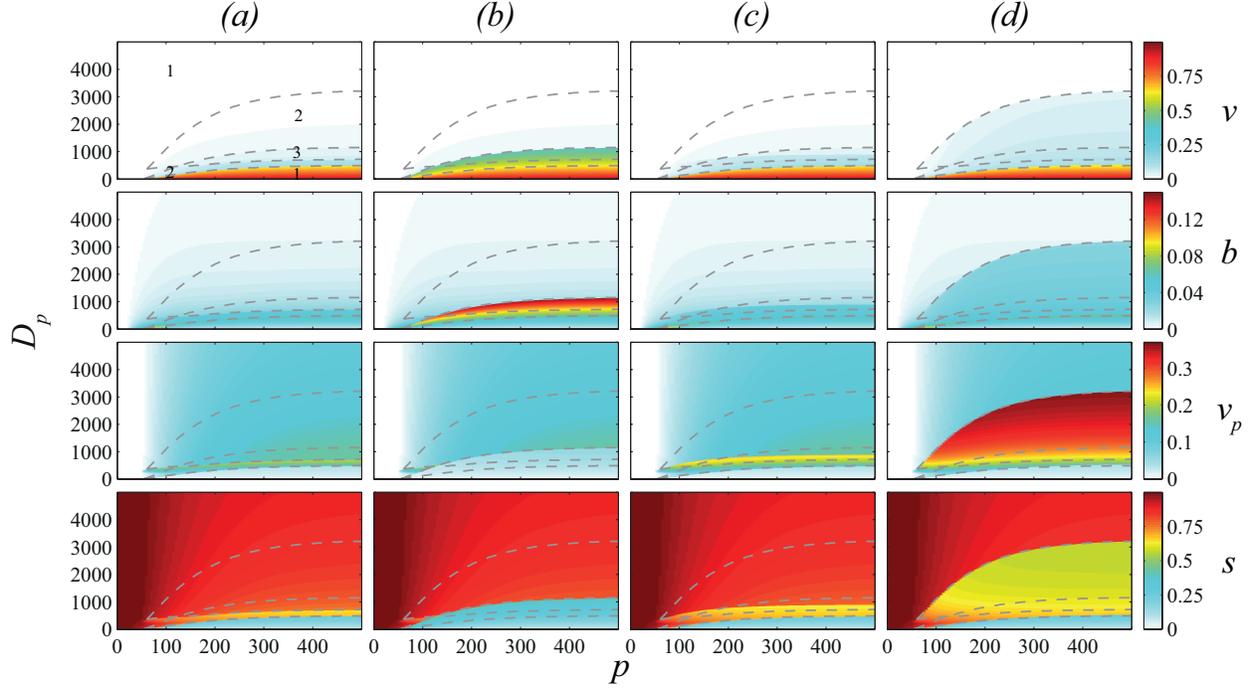}%
\caption{The steady states corresponding to different initial
  conditions as predicted by model I as a function of $p$ and
  $D_p$. Column (a) corresponds to the bare 
  dune initial condition, $v(t=0)=b(t=0)=v_p(t=0)=0$. Column (b)
  corresponds to the vegetation-covered dune initial
  condition, $v(t=0)=1;b(t=0)=v_p(t=0)=0$. Column (c) corresponds to
  the crust-covered dune initial condition,
  $v(t=0)=0;b(t=0)=1;v_p(t=0)=0$. Column (d) corresponds to
  the psammophilous plant-covered dune initial condition,
  $v(t=0)=b(t=0)=0;v_p(t=0)=1$. The different rows correspond to the
  different cover type fractions as indicated. 
  The dashed
  lines mark the edges of the multi-stability regimes, namely,
  the crossing lines between regions with different numbers of physical
  solutions as shown in Fig. \ref{fig:Nsol}(c). The numbers in the top
  left panel indicate the number of stable physical solutions in each
  region.
  \label{fig:ICI}}
\end{figure}

The bifurcation diagrams alone do not elucidate all the information provided by  
the models. The dynamics is of relevance and importance to
understanding the role of psammophilous plants in sand dune
dynamics. We started exploring the dynamics predicted by the models by
investigating the steady state reached from different initial
conditions. In Fig. \ref{fig:ICI}, we show the steady states reached
by different initial conditions calculated using model I. Columns
(a)-(d) correspond to the initial conditions of bare sand dunes ($v(t=0)=b(t=0)=v_p(t=0)=0$),
vegetation-covered sand dunes ($v(t=0)=1;b(t=0)=v_p(t=0)=0$), BSC-
covered sand dunes ($v(t=0)=0;b(t=0)=1;v_p(t=0)=0$) and psammophilous
plant-covered sand dunes ($v(t=0)=b(t=0)=0;v_p(t=0)=1$), respectively. The different rows correspond to the cover type
fractions. Here again, for convenience, we show the exposed sand
fraction. 

The different initial conditions resulted in different steady state
maps. For all initial conditions, we found that for a low drift potential
and a not too low precipitation rate (the lowest part of the panels of
the first row in Fig. \ref{fig:ICI}), the vegetation cover dominates
and stabilizes the sand dunes. For the full vegetation cover initial
condition, the vegetation remains dominant, even at higher values of
the drift potential (see column (b) of Fig. \ref{fig:ICI}). For all
initial conditions and climatic conditions, except for a small regime of
intermediate drift potential and low precipitation, the fraction of
BSC cover is relatively small. The steady state with a maximal fraction
of BSC cover is obtained for intermediate values of the drift
potential and a not too small precipitation rate for an initial
condition of full vegetation cover (see column (b) of
Fig. \ref{fig:ICI}). 

The psammophilous plant cover fraction is significant for
intermediate and high values of the drift potential for all initial
conditions except for the full vegetation cover initial condition for
which $v_p$ only dominates at high values of the drift potential. The
maximal value of $v_p$ is obtained for the $v_p=1$ initial
condition. These results suggest that the basin of attraction of the
steady state solution with $v_p\sim v_c$ is relatively small if there 
is a stable state with a high value of $v$.

The bottom row in Fig. \ref{fig:ICI} shows that for a bare dune
initial condition, stabilization of the dunes is only possible at a high
enough precipitation rate and a low drift potential. For a small range
of intermediate drift potential, the steady state is partially covered
dunes, namely, $v_t\sim v_c$. For the $v=1$ initial condition, we found
that the dunes remain stabilized for a high enough precipitation rate
and an intermediate or weak drift potential. Note that for this initial
condition, the partially covered dune steady state does not appear
(for any climatic condition). For the $b=1$ initial condition the
steady state map is very similar to the map obtained for the bare
dune initial condition. However, the region of partially covered
dunes in steady state is larger and extends to higher values of the
drift potential. For the $v_p=1$ initial condition, the steady state
map shows a large region of partially covered dunes in steady
state. Here, this region extends to very high values of the drift
potential.

\begin{figure}
\includegraphics[width=0.5\linewidth]{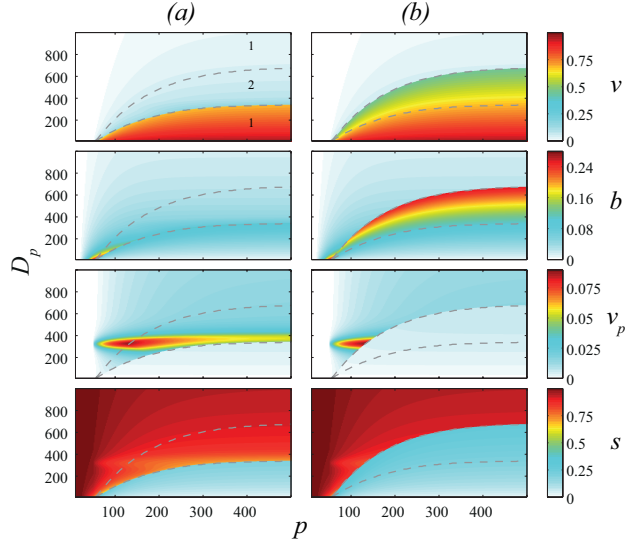}%
\caption{The steady states corresponding to different initial
  conditions as a function of $p$ and $D_p$, as predicted by model
  II. The left column corresponds to 
  the bare dune initial condition, $v(t=0)=b(t=0)=v_p(t=0)=0$, and the right
  column corresponds to the full vegetation cover initial condition,
  $v(t=0)=1;b(t=0)=v_p(t=0)=0$. The different rows correspond to the
  different cover type fractions as indicated. The dashed
  lines mark the edges of the bi-stability regime as shown in Fig. \ref{fig:Nsol}(d).
  The numbers in the top left panel indicate the number of stable
  solutions in each region. 
  \label{fig:ICII}}
\end{figure}

Fig. \ref{fig:ICII} shows the steady states of model II reached by
different initial conditions. Column (a) corresponds to the bare dune
initial condition and column (b) corresponds to the full vegetation cover
initial condition. Initial conditions of full psammophilous plant and
BSC cover resulted in the same steady state as the bare dune initial
condition. These results suggest that the basins of attraction of the
states in the bi-stability regime are determined by the value of $v$
and are less sensitive to the values of $b$ and $v_p$ in this model.
Similarly to model I, we found that in model II, for all initial
conditions, a low drift potential and a not too low precipitation rate, the
vegetation cover dominates and stabilizes the sand dunes. For the full
vegetation cover initial condition, the vegetation remains dominant
even at higher values of the drift potential (see column (b) of
Fig. \ref{fig:ICII}). Note that for the parameters used here, these
values of the drift potential are significantly lower than the values
predicted by model I; it is possible to extend these regions by
choosing a larger maximal growth rate, $\alpha_{v,max}$.
For the bare dune initial condition, the fraction of BSC cover is
small in all climatic conditions except for a small region of low
precipitation and drift potential at which only the BSC can grow. For
the $v=1$ initial condition and not too low values of the drift
potential, the values of $b$ are significant; yet, these values are
smaller than the values of $v$ under the same climatic conditions. We
see that for the parameters used in model II, the maximal values of
$v_p$ are significantly smaller than those obtained for model I. The
only regions with significant values of $v_p$ in steady state are
around ${D_{p,{\rm opt}}}$. For the bare dune initial condition, the
region extends to high precipitation rates, while for the full
vegetation cover initial condition, this region is truncated at low
precipitation rates. The bottom row of Fig. \ref{fig:ICII} shows that
the stability map of the sand dunes is similar to one obtained when
the psammophilous plants are neglected and only the vegetation and BSC
are considered as dynamical variables.

\begin{figure}
\includegraphics[width=0.5\linewidth]{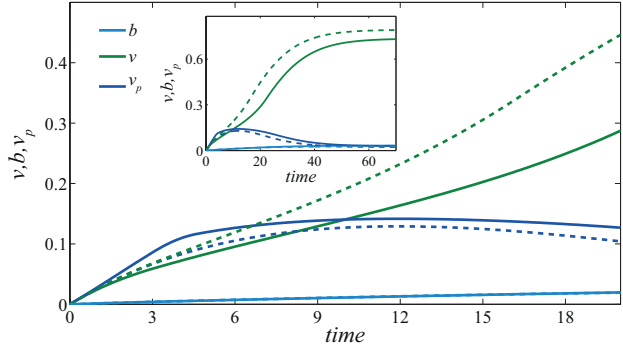}%
\caption{Time evolution of the cover type fractions as predicted by
  model I. The precipitation rate was set to $p=300$. The dashed lines
  correspond to drift potential, $D_p=200$, and the solid lines correspond
  to $D_p=300$. The solid lines show dynamics in which the
  psammophilous plants initially dominate, and later on, the
  normal vegetation dominates. The dashed lines show an evolution in
  which the psammophilous plants and the normal vegetation grow
  equally at first, and later on, the normal vegetation dominates. 
  \label{fig:dynamics}}
\end{figure}

A common paradigm for the stabilization of sand dunes under high sand
drift is that the psammophilous plants act as pioneers
\cite{Danin-1996:plants}. Due to their ability to flourish under
significant sand drift, they are the first to colonize bare
dunes. Their growth reduces the sand drift by wind shading and enables
the growth of regular vegetation, eventually resulting in stabilized
dunes in which the fraction of vegetation cover dominates. In order to
test if our models are capable of reproducing this paradigm, we
investigated the temporal dynamics. In Fig. \ref{fig:dynamics}, we show
the values of $v$, $b$ and $v_p$ versus the time for the bare dune initial
condition. The precipitation rate was set to $p=300$. The dashed lines
correspond to $D_p=200$, and the solid lines correspond to
$D_p=300$. Our results show that under some climatic conditions, the
dynamics follows the paradigm (the solid lines), while under other
climatic conditions, the initial growth of the vegetation is identical
to the initial growth of the psammophilous plants (the dashed
lines). The results presented in Fig. \ref{fig:dynamics} were
calculated using model I. For the same climatic conditions, model II
yields qualitatively the same results.
\section{Summary and discussion}
We have studied the effect of psammophilous plants on the dynamics of
sand dunes using a simple mean field model for sand dune cover
dynamics (vegetation, BSC and psammophilous plants). Two main
mechanisms of interaction of the psammophilous plants with the wind
and the sand drift were modeled separately. Root exposure and
the covering of branches/leaves by the sand drift enhance the net growth of
 psammophilous plants and result in maximal growth under optimal
sand drift conditions (model I). Branches/leaves torn off by the wind and buried
by the sand develop new plants and increase the growth rate
of psammophilous plants under an optimal wind drift potential
(model II).  These two approaches resulted in qualitatively different
steady states and dynamics; the sand drift optimal growth model (model
I) shows a richer steady state map, with up to three stable dune
states (extensive sand cover, moderate sand cover and small sand
cover) while the wind drift potential optimal growth rate (model II)
shows up to two stable dune states (extensive and small sand cover).

While there are examples for the coexistence of active and fixed dunes
under similar climate conditions
\cite[][]{Yizhaq-Ashkenazy-Tsoar-2007:why,Yizhaq-Ashkenazy-Tsoar-2009:sand},
we are not aware of observations that may be associated with the
``new'' dune state predicted by model I of moderate cover in which the
vegetation cover is close to the critical vegetation cover, $v_c$ (it
is important to note that according to this model, the basin of
attraction of this state is very small, and therefore, it may not be
easily realized in observations). Identifying such a state in
observations will provide strong support for the model's setup. We
hope to explore this and other features of the model in the future.

The model proposed here does not aim to be operative. 
It aims to provide a qualitative understanding of the dynamics 
of psammophilous plants in the presence of regular
vegetation and BSC. Nevertheless, a comparison with observations of
the bi-stability of sand dunes reported in
\cite{Yizhaq-Ashkenazy-Tsoar-2007:why,Yizhaq-Ashkenazy-Tsoar-2009:sand}
indicates that model I fits the observations better than
model II; model II exhibits only an active dune state for drift
potential values that are larger than 700 (see Fig. \ref{fig:Nsol}),
while stable dunes exist in nature for higher values of $D_p$. Model II can to be tuned
to fit these observations by increasing $\alpha\_max_{v}$ or
decreasing $\epsilon_v$. We do not present these results here, in order to use the same parameters in models I and II wherever possible. Another observation that is worth
noting concerns the fraction of BSC cover. Both model I and II resulted in BSC cover
that does not exceed $25\%$. Studies have reported almost complete BSC cover
on sand dunes for small plots (order of meters) \cite[e.g., Fig. 5B
and 7 of][]{Veste-Breckle-Eggert-Littmann-2011:vegetation}.
This discrepancy between the model and observations can be attributed
to the mean field nature of the model, representing only scales of $kms$ that
usually contain several types of dune surface covers. 

The separation of the two growth mechanisms--by assuming optimal
growth conditions under either (i) an optimal sand drift or (ii) an
optimal wind drift potential--helps us to understand the effect of each of
these mechanisms. Yet, a more realistic model should include a
combination of these two mechanisms, resulting in, most probably,
a richer dynamics and steady state map. Observations regarding
psammophilous plants may help to determine the role of each mechanism,
which one is more favorable, and if a combination of the two is
plausible.

For the parameters used here, model I predicts that the psammophilous
plant cover can reach the critical value for sand-drift shading, while
model II predicts that their fraction of cover is very small.
According to both models, there are climate conditions under which the
steady state picture is not affected by the presence of the
psammophilous plant cover fraction as an additional dynamical
variable in the model. For example, under a low wind drift potential
and a high precipitation rate, regular vegetation is the dominant cover
type, and the BSC and the psammophilous plants may be ignored. Under
extremely dry conditions, $p<50mm/yr$, and a low wind drift potential,
the BSC will be the dominant cover type, and both regular vegetation
and psammophilous plants may be ignored. Under an extremely high wind
drift potential, the dunes will be fully active without any surface
cover. Psammophilous plants may be the dominant cover type under a high
enough precipitation rate ($p>50mm/yr$) and a strong wind drift
potential (the definition of strong depends on the model (I or II) and
the parameters).

While some of the cover types can be ignored in the steady state, they
can still greatly influence the dynamics leading to the observed
steady state. We have demonstrated that starting from a bare dune
state, psammophilous plants may be the first to grow, reducing the sand
drift and thus enabling the growth of regular vegetation which
eventually dominates and stabilizes the dunes
\cite[][]{Danin-1996:plants}. This, however, is not always the case as
our model predicts that under different climate conditions (wind drift
potential), the regular vegetation and psammophilous plants grow
equally at first, cooperating in reducing the sand drift, followed by
a faster growth of the regular vegetation, which eventually dominates and
stabilizes the dunes.

Our preliminary numerical results suggest the possibility of a Hopf
bifurcation leading to oscillatory behavior of the different cover
types. This behavior and its relevance to observations, as well as
the incorporation of spatial effects in the model, as was done in
\cite{Yizhaq-Ashkenazy-Levin-Tsoar-2013:spatiotemporal}, are left for
future research. In addition, we plan to use this model and its
extensions to study the response of sand dunes to different scenarios
of climate change.

\begin{acknowledgments}
  The research leading to these results has received funding from the
  European Union Seventh Framework Programme (FP7/2007-2013) under
  grant number 293825. This research was also supported by the Israel
  Science Foundation. We thank Dotan Perlstein for his contribution 
  during the first stages of this research and Shai Kinast for 
  helpful discussions.
\end{acknowledgments}

\bibliography{psamdyn}

\end{document}